\documentclass[conference]{IEEEtran}
\usepackage{amsmath,amssymb,amsfonts}
\usepackage{algorithmic}
\usepackage{graphicx}
\usepackage{textcomp}
\usepackage{xcolor}
\usepackage{url}
\usepackage[caption=false]{subfig}
\usepackage{hyperref}
\usepackage{filecontents}
\usepackage[noadjust]{cite}

\begin{document}

\title{Modeling the Epidemic Outbreak and Dynamics of COVID-19 in Croatia}
\author{\IEEEauthorblockN{Ante Loji\'{c} Kapetanovi\'{c}\IEEEauthorrefmark{1}, Dragan Poljak\IEEEauthorrefmark{2}}
\IEEEauthorblockA{\textit{Faculty of Electrical Engineering, Mechanical Engineering and Naval Architecture}\\
\textit{University of Split}\\
Split, Croatia \\
\IEEEauthorrefmark{1}alojic00@fesb.hr,
\IEEEauthorrefmark{2}dpoljak@fesb.hr}}

\maketitle

\begin{abstract}
\label{sec.abstract}
The paper deals with a modeling of the ongoing epidemic caused by Coronavirus disease 2019 (COVID-19) on the closed territory of the Republic of Croatia. Using the official public information on the number of confirmed infected, recovered and deceased individuals, the modified SEIR compartmental model is developed to describe the underlying dynamics of the epidemic. Fitted modified SEIR model provides the prediction of the disease progression in the near future, considering strict control interventions by means of social distancing and quarantine for infected and at-risk individuals introduced at the beginning of COVID-19 spread on February, 25\textsuperscript{th} by Croatian Ministry of Health. Assuming the accuracy of provided data and satisfactory representativeness of the model used, the basic reproduction number is derived. Obtained results portray potential positive developments and justify the stringent precautionary measures introduced by the Ministry of Health.
\end{abstract}
 

\begin{figure*}[htb!]
    \centering
    \includegraphics[width=0.9\textwidth]{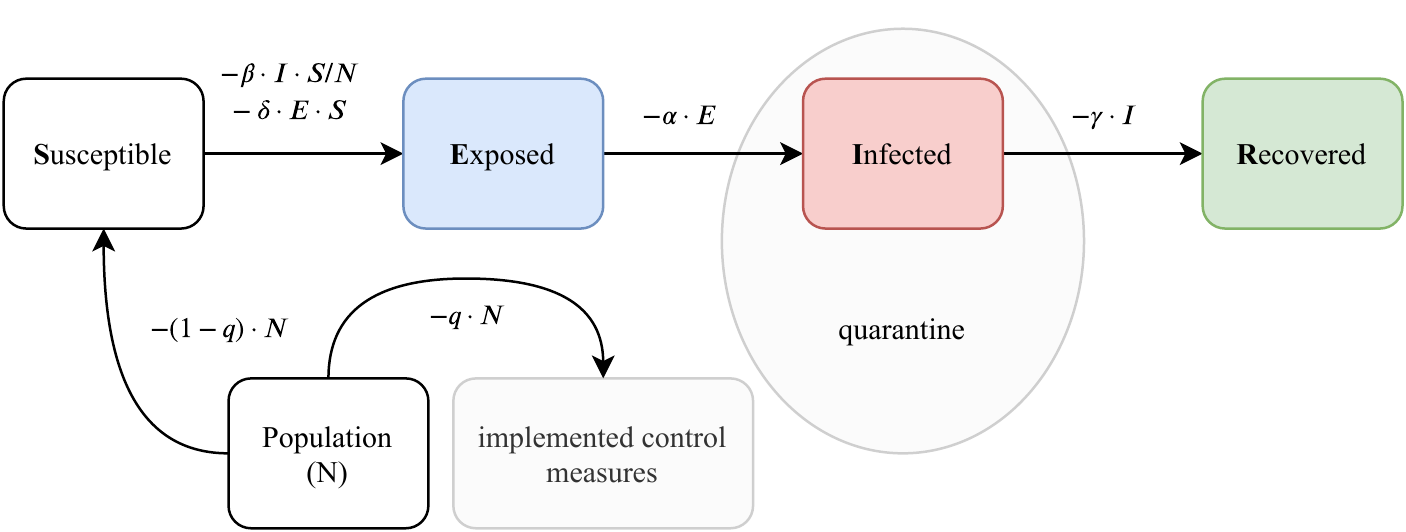}
    \caption{The modified SEIR compartmental model of COVID-19 disease.}
    \label{fig.seir}
\end{figure*}

\section{Introduction}
\label{sec.introduction}
Coronavirus 2019-2020 epidemic of coronavirus disase 2019 (COVID-19), caused by severe acute respiratory syndrome coronavirus 2 (SARS-CoV-2), began in Wuhan, China, in late December 2019 \cite{who_covid_2020}. Although Europe had time to prepare for the case of an epidemic outbreak, the lack of relevant medical data and the novelty of COVID-19 disease have made it the new epicenter of the current pandemic with more reported cases and deaths than the rest of the world combined, excluding China, at the time of writing this paper. 

The paper aims to model the disease dynamics by determining epidemic parameters in order to estimate the number of infected individuals over time on the closed territory of Croatia and to predict the pace of COVID-19 disease spreading by using the basic reproduction number $R_0$. Since control measures in Croatia, such as tracing close contacts, quarantining infected cases, promoting social distancing and self-protection using protective equipment, i.e. medical masks and gloves in a public area, started with the first confirmed case on February, 25th, $R_0$ explicitly indicates the effectiveness of the current control measures and facilitate the adoption and implementation of potential additional measures. Thus, the ultimate goal of this paper is to accurately determine $R_0$ as the expected number of secondary infections caused by a single infectious symptomatic or asymptomatic individual in a fully susceptible population \cite{Kretzschmar2016}.

Even though Coronavirus-based diseases are well documented and modeled with a satisfactory level of accuracy in computational epidemiology literature \cite{Han2009}, there are a few novel properties of COVID-19 disease that need to be considered. Firstly, a long incubation period will cause time delay between real dynamics and the actual status. Furthermore, both symptomatic and asymptomatic individuals are capable of being infectious carriers of SARS-CoV-2. Finally, the disease transmission is achieved via respiratory droplets and is extremely difficult to prevent due the well resilient pathogen hardly affected by external atmospheric conditions \cite{Guan2020}. 

In order to determine epidemic parameters, stochastic models are well adopted and preferred in the current research \cite{Marathe2020}, but during the initial stage of the epidemic, the data are sparse and the epidemic dynamics are better described using deterministic modeling. Generalizations of the Susceptible-Exposed-Infected-Recovered (SEIR) model are used for China \cite{Peng2020,Zhao2020} and heavily affected parts of Europe \cite{Cereda2020,Lopez2020}. However, due to the change of diagnostics and the lack of implementation of control measures in the initial stage of the outbreak, early published models were forced to add many free parameters and fit the data in multiple phases. Such complicated models are prone to over-fitting. 

This paper introduces the modified version of SEIR model with a single additional parameter to fit and implicitly included additional compartment for quarantined and self-isolated individuals. The paper is organized as follows: Section \ref{sec.modeling} outlines the mathematical modeling of COVID-19 and methods for predicting the spread of the mentioned disease; Section \ref{sec.results} is dedicated to the computational results and discussion; lastly, conclusion is presented in Section \ref{sec.conclusion}. Scraping of official data, the modified SEIR model fitting details and optimization method are given in Appendix \ref{sec.mathdetails}, while additional details on epidemic parameters are given in Appendix \ref{sec.params}. 

\section{Tools for Mathematical Modeling of COVID-19}
\label{sec.modeling}
\subsection{The Modified SEIR Compartmental Model}
\label{sec.modeling.seir}
SEIR model, based on the early work of Kermack and McKendrick \cite{Kermack1991_1,Kermack1991_2,Kermack1991_3}, is compartmental, population-based epidemiological model used for an appropriate mathematical representation of underlying dynamics of an infectious disease. Such model, readily generalized and computationally tractable is a perfect candidate to simulate a propagation of a disease without much of a prior knowledge on a disease. 
The entire population, which is assumed to be constant in time (no vital dynamics by means of birth and death rate are taken into an account) is divided into 4 compartments. The rate of change over time for each compartment is expressed in terms of a set of 4 coupled ordinary differential equations (ODEs). 
Note that the nature of differential equations dealing with time dependant quantities is to assume continuous variables, which is not the case for both the number of people and observed time steps. 
In order to obtain the discrete time equivalent, the changes of individuals in each compartment are modeled via a set of difference equations over an arbitrary discrete-time step $t$ (typically a day). It is worth emphasizing that the discretization of variables is usually carried out as a prequel to stochastic modeling \cite{canto_2015}. Nevertheless, continuous compartmental models provide more detailed epidemic dynamics \cite{Peng2020}, thus the modified SEIR model is defined in terms of the following set of ODEs:
\begin{align}
    \label{eqn.s}
    \frac{dS}{dt} &= - \beta \cdot \frac{I}{N} \cdot S - \delta \cdot E \cdot S \\
    \label{eqn.e}
    \frac{dE}{dt} &= \beta \cdot \frac{I}{N} \cdot S - \alpha \cdot E + \delta \cdot E \cdot S \\
    \label{eqn.i}
    \frac{dI}{dt} &= \alpha \cdot E - \gamma \cdot I \\
    \label{eqn.r}
    \frac{dR}{dt} &= \gamma \cdot I
\end{align}
where,
\begin{itemize}
    \item $S$ is the susceptibles compartment;
    \item $E$ is the exposed compartment;
    \item $I$ is the infected compartment and
    \item $R$ is the recovered (or deceased) compartment.
\end{itemize}
A simplified visual representation of the modified SEIR model of COVID-19 disease based on the set of equations (\ref{eqn.s})-(\ref{eqn.r}) is shown in Fig. \ref{fig.seir}. It is assumed that only a small portion of the total population $N$ is susceptible to the infection due to timely and effective measures implemented from the first day of the epidemic.

Free parameters are in order: the transition rate, $\beta$, which stands for the average number of contacts per unit time multiplied by the probability of transmission per each contact, the direct transition rate between exposed and susceptible compartment, $\delta$; the reciprocal value of the incubation period, $\alpha$, and the rate of recovery or mortality, $\gamma$.
The mentioned expansion with respect to standard SEIR model lies in the expression $\delta \cdot E \cdot S$,  which allows exposed individuals, considered to be in the incubation phase, to transmit the infection onto susceptible individuals \cite{Wei2020}. 

\subsection{$R_0$ - The Key Epidemic Parameter}
\label{sec.modeling.R_0}
$R_0$ (read ‘R naught’) is the expected number of secondary infections in a sufficiently large population without prior immunity to a disease. The assumption is well aligned with the current COVID-19 disease outbreak, since there is no maternal immunity nor there is a functional vaccine yet.
$R_0$ number depends on both biological characteristics of the pathogen and the host, and on social circumstances, as follows: 
\begin{equation} \label{eqn.R_0}
    R_0 = \frac{\beta}{\gamma + \alpha}
\end{equation}{}
The importance of $R_0$ lies in the fact that it represents somewhat of a threshold value from which one could determine the further progress of the disease. Namely, if $R_0 > 1$, the pathogen is expected to invade a population - the epidemic expansion is likely to occur. On the other hand, if $R_0 < 1$, the ongoing disease will vanish over time. The goal of preventive measures is to keep $R_0$ below the mentioned threshold.
In other words, $R_0$ directly depends on the total infectious period,
\begin{equation}
    \tau = \frac{1}{\gamma + \alpha}
\end{equation}
which is considered to be constant biological property and on the probability of disease transmission per contact per unit time, $\beta$. The transition rate, $\beta$, can be implicitly lowered by lowering the susceptible population through social distancing and by lowering the transition rate, e.g. enhanced hygiene, wearing protective clothing. 

In this paper, $R_0$ is determined via expression given in (\ref{eqn.R_0}), with parameters obtained through fitting the actual data, time series of confirmed active infections and recovered individuals from February, 25\textsuperscript{th} to April, 11\textsuperscript{th}, to the modified SEIR model. The data are obtained from the official publicly available sources \cite{koronavirushr}.

\begin{figure}[]
    \centering
    \includegraphics[width=\linewidth]{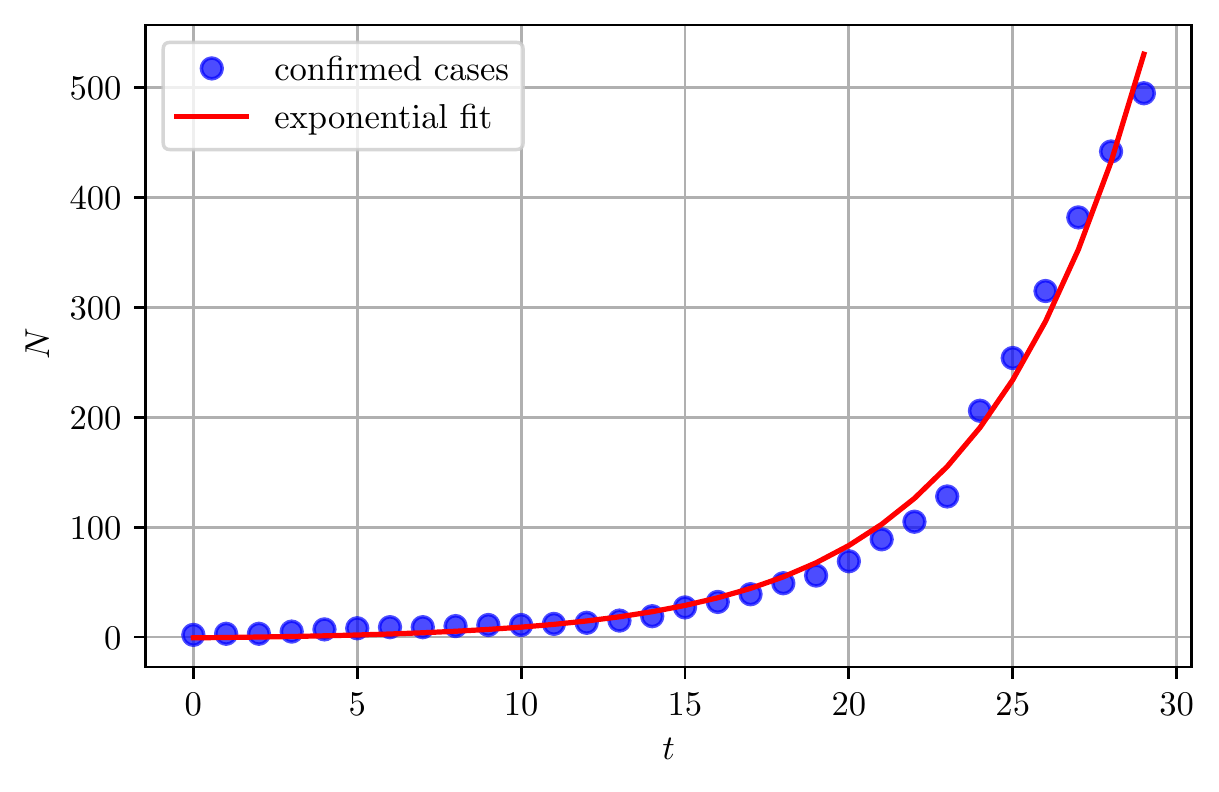}
    \caption{Exponential function fitted to the number of confirmed infected cases, $N$, during the first $t=30$ days of COVID-19 disease outbreak in Croatia. Even though, control measures were implemented from the first day of the disease, because of the incubation period of 3-15 days, the effect of those measures has been shown with a time lag.}
    \label{fig.early_exp}
\end{figure}

\section{Results and Discussion}
\label{sec.results}
An early stage of the epidemic spread of an infectious disease, shown in Fig. \ref{fig.early_exp}, is often characterized by exponential growth, with the rapidity of growth dependent on the force of infection, formally defined as the per capita rate at which susceptible individuals contract infection and mathematically formulated with the following expression:
\begin{equation}
\label{eqn.force}
    F = \beta \cdot I
\end{equation}

\begin{figure}[!t]
    \centering
    \subfloat[Exponential function fitted to the confirmed number of infected individuals $N$ over time $t$. The dashed line represents prediction for the upcoming 5 days.]{\includegraphics[clip,width=\columnwidth]{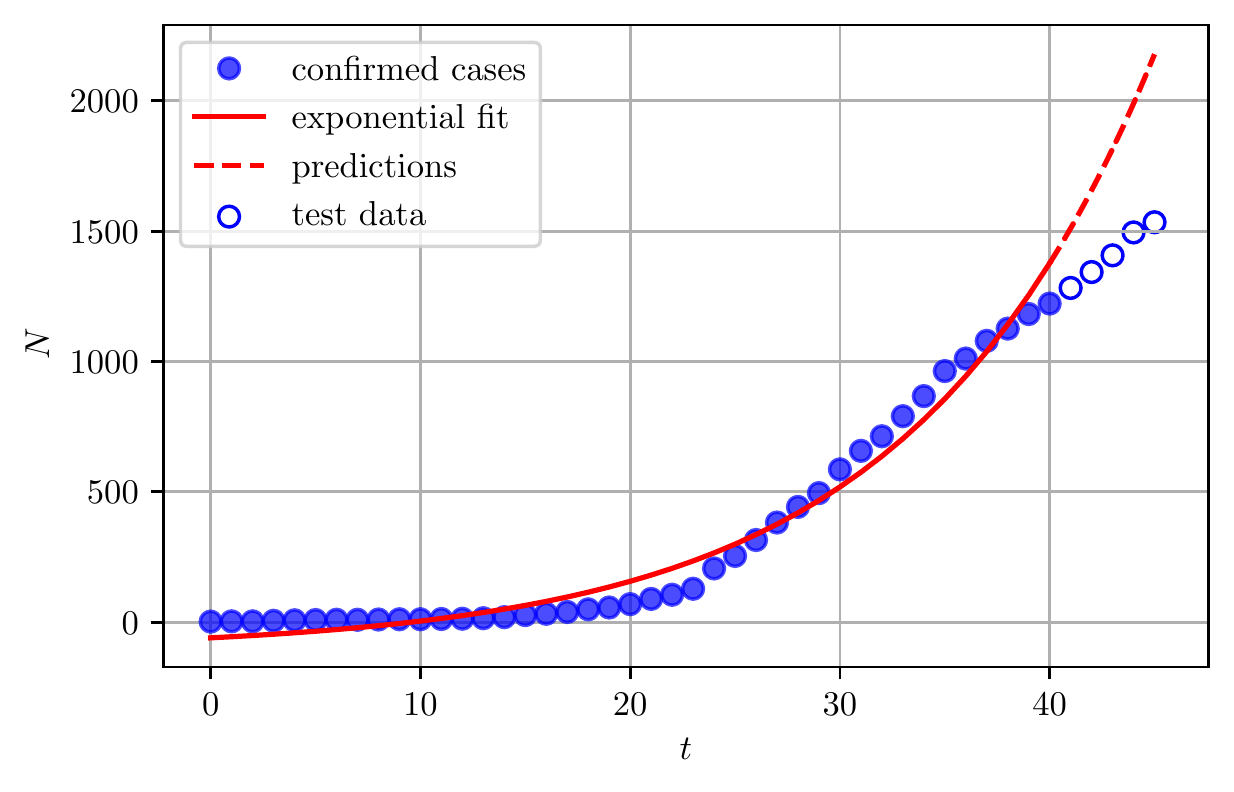}
    \label{fig.exp}}
    \hfil
    \subfloat[Logistic function fitted to the confirmed number of infected individuals $N$ over time $t$. The dashed line represents prediction for the upcoming 5 days.]{\includegraphics[clip,width=\columnwidth]{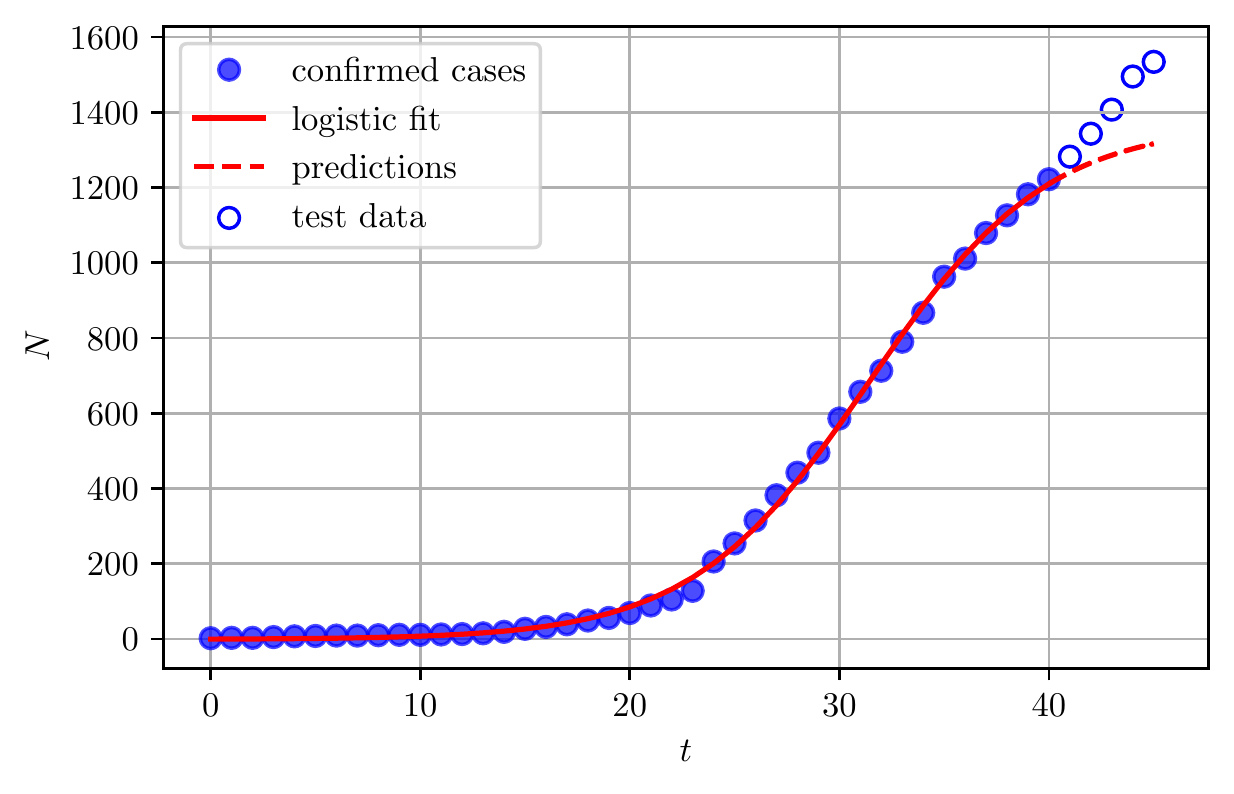}
    \label{fig.logit}}
    \caption{Exponential and logistic functions fitted to the data divided into training and testing set.}
    \label{fig.exp-logit}
\end{figure}

After an early stage, exponential growth is physically unfeasible because of the finite size of the population. Exponential function overestimates the total number of infected individuals, shown in Fig. \ref{fig.exp}, where actual growth soon starts to asymptotically approach the value of the maximum infections and is more accurately represented via the logistic function. However, since the current status of the growth is unknown, the logistic function fitted to the portion of the data provides grossly underestimated predictions, depicted in Fig. \ref{fig.logit}.

\begin{figure}[]
    \centering
    \includegraphics[width=\linewidth]{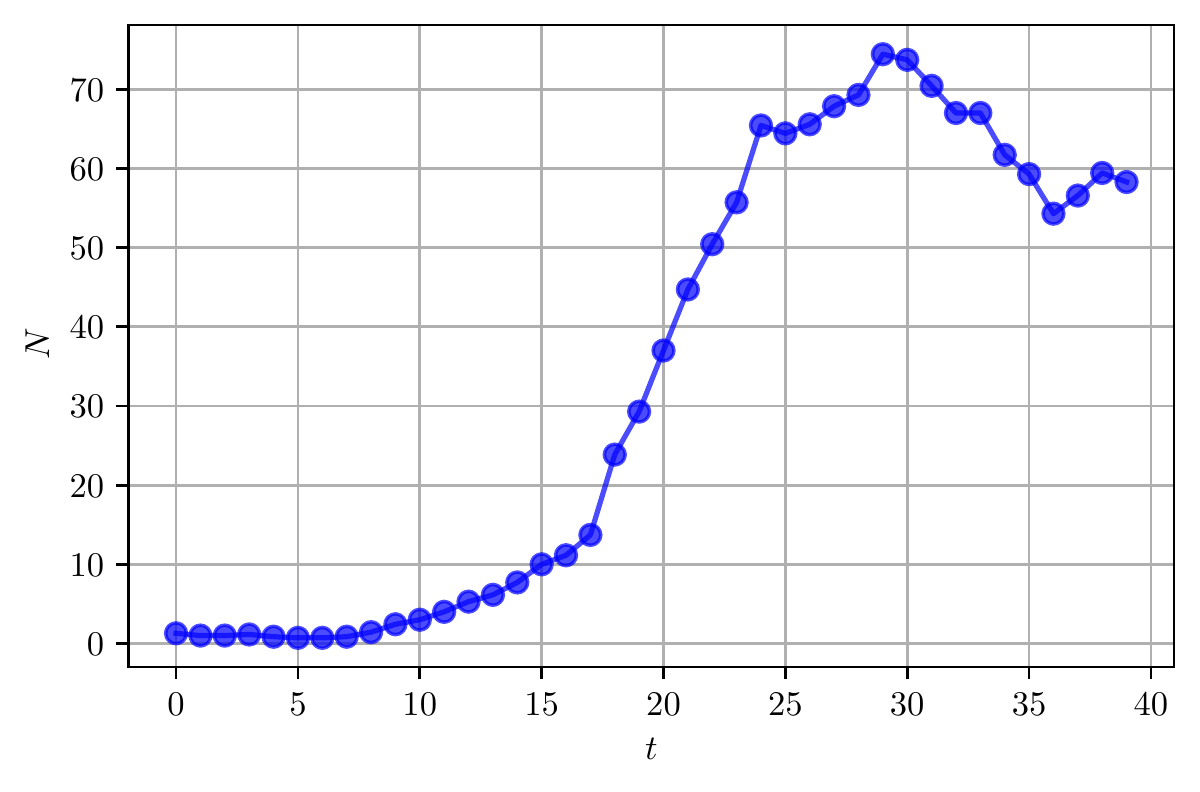}
    \caption{Time series of new confirmed cases $N$ averaged over 7-day period.}
    \label{fig.newcases}
\end{figure}

\begin{figure}[]
    \centering
    \includegraphics[width=\linewidth]{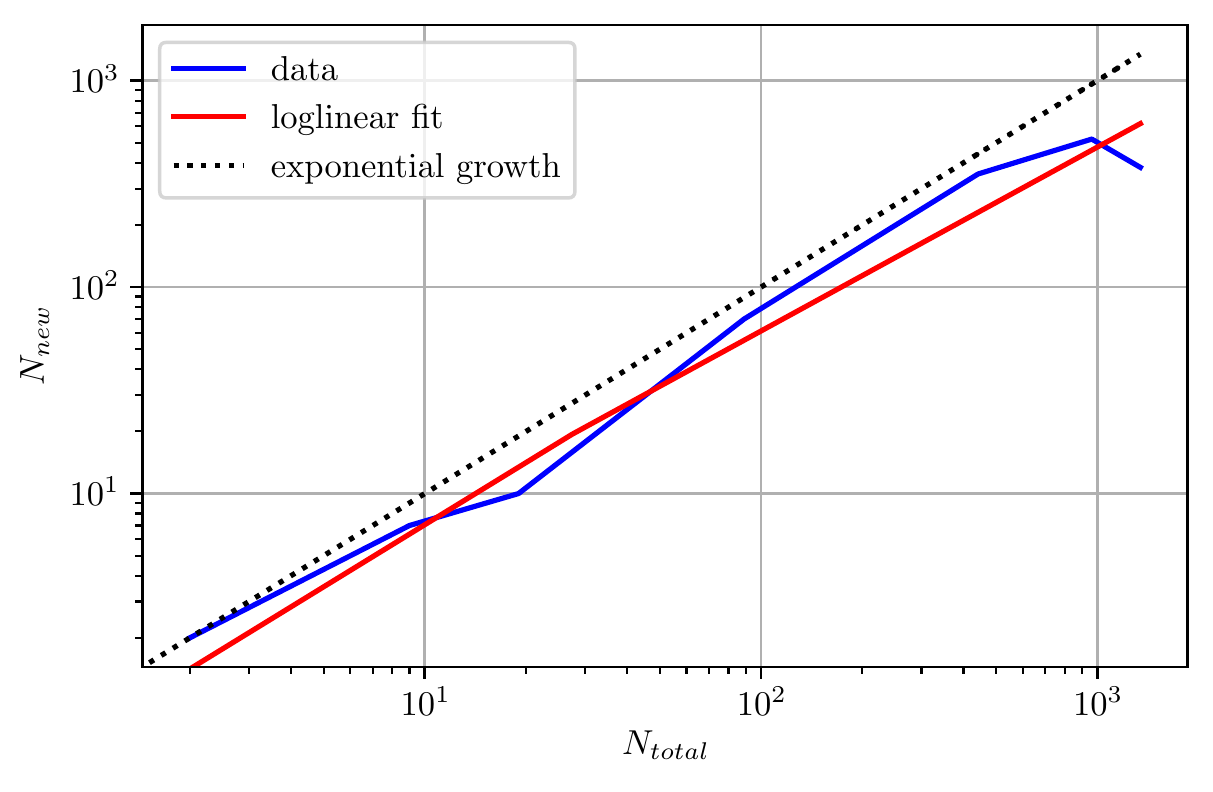}
    \caption{Logarithmic number of new cases, $N_{new}$, vs. logarithmic number of total confirmed cases, $N_{total}$, averaged over 7-day period (blue line) compared to the theoretical exponential function (black dashed line) in log-scale.}
    \label{fig.new-v-total}
\end{figure}

Instead of observing confirmed cases over time, a more appropriate capture of the current trend of the disease spread is achieved by observing the change itself. Fig. \ref{fig.newcases} charts the number of new cases averaged over previous 7 days. Obviously, COVID-19 disease spread in Croatia is in a downward trend. 
If the influence of time is instead taken implicitly and the growth rate is plotted against the current number of infections in logarithmic scale, the resulting comparison between the actual growth of COVID-19 disease and the exponential growth will be easier to compare, see Fig. \ref{fig.new-v-total}. Once the trend begins to steeply decline, the epidemic is considered to be terminated, i.e. the reproductive number drops below 1. 

\begin{figure}[!t]
    \centering
    \subfloat[]{\includegraphics[clip,width=\columnwidth]{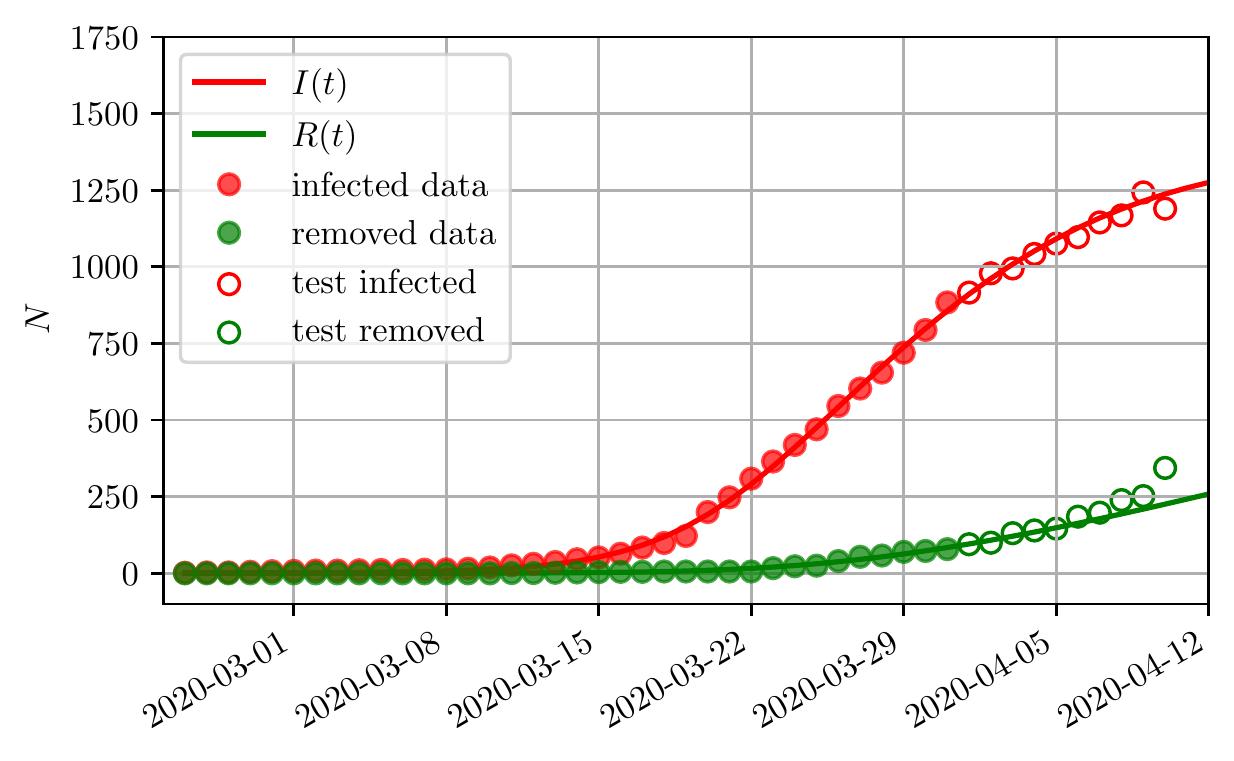}
    \label{fig.seiR_0.80}}
    \hfil
    
    \subfloat[]{\includegraphics[clip,width=\columnwidth]{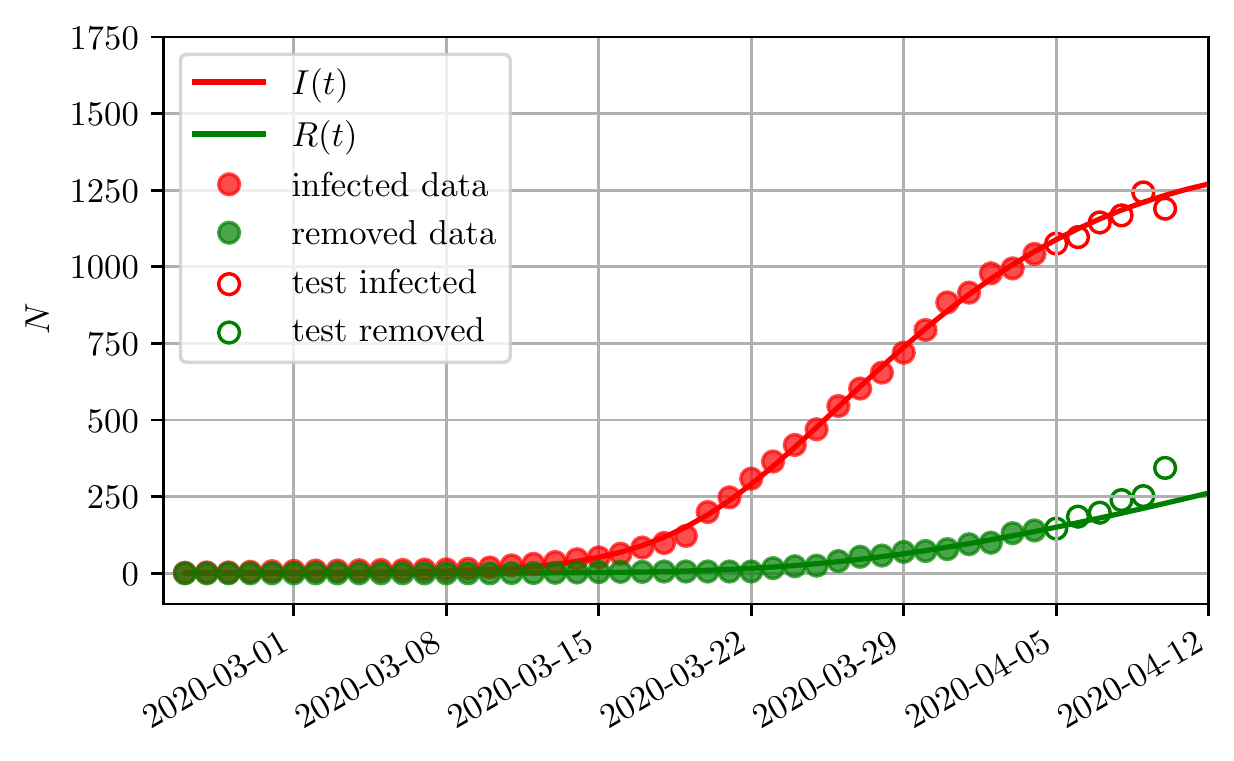}
    \label{fig.seiR_0.88}}
    \hfil
    
    \subfloat[]{\includegraphics[clip,width=\columnwidth]{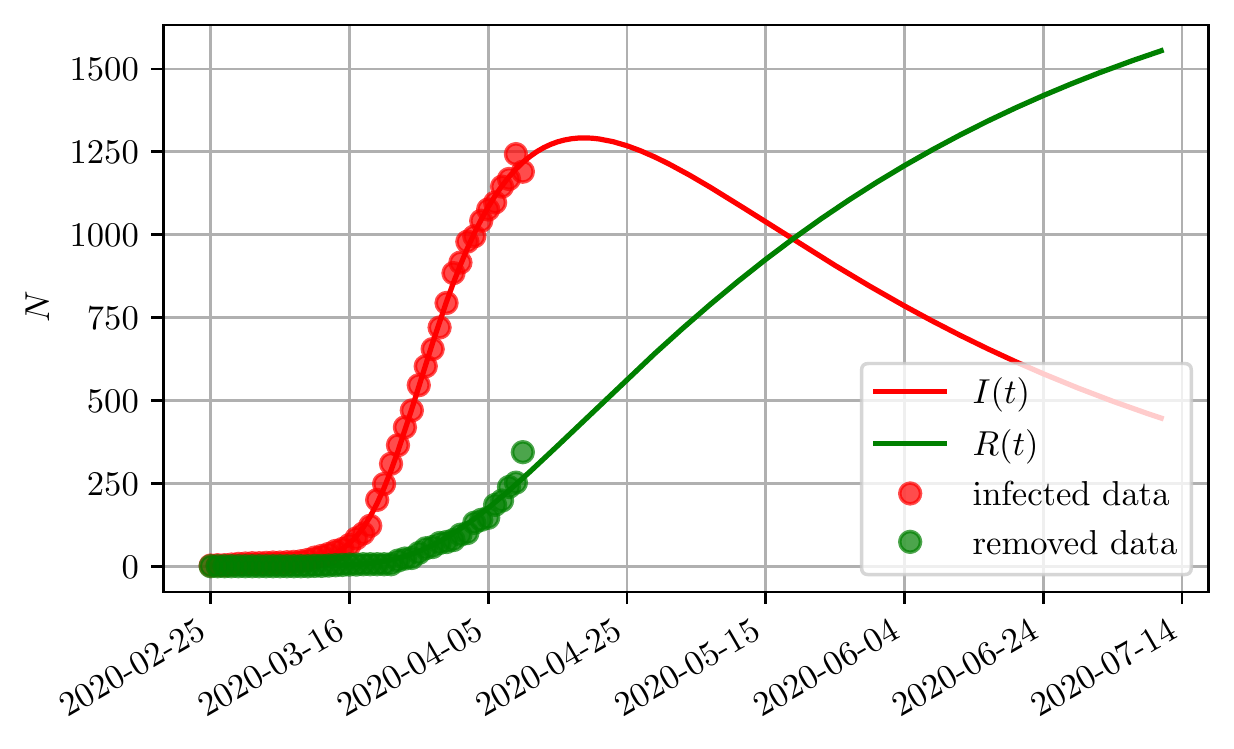}
    \label{fig.seir1}}
    
    \caption{The modified SEIR model fitted to different sizes of the training data set. The top two figures compare the actual data with the predicted solution of the infected and the recovered (or deceases) differential equations defined in (\ref{eqn.i}) and (\ref{eqn.r}) respectively. The bottom figure shows the prediction of the near future.}
    \label{fig.seirfit}
\end{figure}

The modified SEIR model, introduced in the section \ref{sec.modeling.seir}, is fitted to the number of confirmed active, recovered and death cases over time. Using the estimated parameters of the fitted model, described in detail in Appendix \ref{sec.params}, reproduction number is calculated for different phases of the epidemic. The data are fitted using the iterative Limited-memory Broyden–Fletcher–Goldfarb–Shanno (L-BFGS) algorithm. Within this framework, for each optimization iteration, the set of ODEs (\ref{eqn.s})-(\ref{eqn.r}) is solved by the 4\textsuperscript{th} order Runge-Kutta method. The optimization procedure and the solution of the set of ODEs are presented in more detail in section \ref{sec.mathdetails}. Fig. \ref{fig.seirfit} depicts fitted modified SEIR model using different sizes of the train data set. Fig. \ref{fig.seiR_0.80} shows the model fitted using 80\% of the data set, while the rest of the data are used in order to assess goodness of fit using the out-of-sample data, Fig. \ref{fig.seiR_0.88} shows the SEIR model fitted using 88\% of the data set, while Fig. \ref{fig.seir1} deals with the complete data set.
Fitting the data provides optimal values for epidemiological parameters $\alpha$, $\beta$, $\gamma$ and $\delta$. $R_0$ is then calculated for different phases of the epidemic and the resulting values are 1.43, 1.33 and 1.25 for 80\% of the data, 88\% of the data and the complete data set, respectively. 
These results imply the effectiveness of the control measures implemented to combat the epidemic as $R_0$ decreases with each increase of the data set. In case there is no change in control measures, one could infer that the positive downward trend will continue up until late April when the number of confirmed active infected cases will reach its maximum. The maximum point is also the inflection point indicating the moment at which $R_0<1$ and after which, with the retention of the control measures, the number of total confirmed cases stops increasing.

\section{Conclusion}
\label{sec.conclusion}
In this paper, the slightly modified SEIR compartmental epidemiological model, regarded as an opener to the subject, is proposed in order to analyze and predict COVID-19 epidemic in Croatia. Transmission parameters of the model are optimized using the available data on the number of confirmed, recovered and deceased individuals, respectively. The model assumes a sufficient percentage of a population $q \cdot N$ being voluntarily self-isolated, shown in Fig. \ref{fig.seir}. The percentages of self-isolated people, together with initial number of exposed, infected and recovered individuals are taken as initial values in order to solve the set of ODEs.

The parameters are used to calculate $R_0$ for different stages of the spread of COVID-19 in Croatia. The values of $R_0$ improve as the data approach closer to April 11\textsuperscript{th}. At this point one considers the government control measures, implemented to slow down the disease spread, to be effective and should be continued until $R_0$ number decreases and remains sufficiently long period of time below the threshold value of 1.

Based on fitted SEIR model, the climax of the infection is expected to occur in late April. The inflection point should follow soon after that date, which ultimately means that the epidemic is in a strong decline period.

\section*{Acknowledgment}
This research has been funded by DATACROSS project of The Centre of Research Excellence for Data Science and Advanced Cooperative Systems (CRE ACROSS-DataScience).

\appendices
\renewcommand{\theequation}{\thesection.\arabic{equation}}
\setcounter{equation}{0}

\section{Outline of Mathematical Modeling}
\label{sec.mathdetails}
The number of confirmed, recovered and deceased individuals is collected from available data for the period of February, 25\textsuperscript{th} to April, 11\textsuperscript{th}. The modified SEIR model consists of 4 free parameters $\beta$, $\gamma$, $\alpha$ and $\delta$ to be optimally determined by minimizing the loss function $J(\beta, \gamma, \alpha, \delta)$. The loss function is formulated as the weighted summation of $l_2$ norm between the actual number of actively inffected, recovered and deceased people and their respective approximations. The loss function is written using the following expression:
\begin{equation}
    J(\beta, \gamma, \alpha, \delta) = \lvert\lvert I(t) - \hat{I}(t)\lvert\lvert_2 + \lvert\lvert R(t) - \hat{R}(t)\lvert\lvert_2
\end{equation}
The approximations are obtained by solving the set of deterministic ODEs and are marked with $\hat{I}(t)$ for approximated number of active infections over time and $\hat{R}(t)$ for approximated total sum of recovered and deceased individuals over time. 

The minimization is performed using L-BFGS method where for each set of parameter values, the 4\textsuperscript{th} order Runge-Kutta method is employed to solve the set of ODEs with known initial values. For each of the differential equations given in (\ref{eqn.s})-(\ref{eqn.r}), initial values are determined using available data on the number of susceptible individuals $S_0$, the number of exposed individuals $E_0$, the number of infected individuals $I_0$ and the number of recovered and deceased individuals $R_0$ at the first day of the epidemic. Both, the minimization and the initial value problem for the set of ODEs are solved using SciPy, open-source scientific software built on the NumPy extension of Python programming language.

\setcounter{equation}{0}

\section{List of Parameters}
\label{sec.params}
\begin{itemize}
    \item $\beta$ - transition or infectious rate; controls the rate of spread which represents the probability of transmitting disease between a susceptible and an infected individual per contact per unit time;
    \item $\gamma$ - recovery or mortality rate is determined by the average duration of an infection, which eventually lead to recovery or death, where individual is no longer classified as susceptible; 
    \item $\alpha$ - incubation rate; the reciprocal value of the incubation period. In the case of COVID-19 disease, an exposed individual is infectious even in the incubation period;
    \item $\delta$ - direct transition rate between susceptible and exposed individual. As mentioned, an exposed individual is capable of shedding the infection even in the asymptomatic phase;
    \item $q$ - quarantine or self-isolation rate; the percentage of the total population insusceptible to COVID-19 disease.
\end{itemize}

\bibliographystyle{IEEEtran}
\bibliography{./main}{}

\end{document}